\documentclass[prd,twocolumn,preprintnumbers,amsmath,amssymb,superscriptaddress,showpacs,showkeys,nofootinbib]{revtex4-1}
\usepackage{amsmath,amsfonts,bm,mathtools}
\usepackage{graphicx}
\usepackage{amsmath}
\usepackage[colorlinks, linkcolor={red},citecolor={blue}]{hyperref}

\begin{document}
\title{Isotropization and change of complexity by gravitational decoupling}
\author{R.~Casadio}
\email{casadio@bo.infn.it}
\affiliation{Dipartimento di Fisica e Astronomia, Alma Mater Universit\`a di Bologna, via Irnerio~46, 40126 Bologna, Italy}
\affiliation{Istituto Nazionale di Fisica Nucleare, Sezione di Bologna, I.S.~FLAG  viale Berti~Pichat~6/2, 40127 Bologna, Italy}
\author{E.~Contreras}
\email{econtreras@yachaytech.edu.ec}
\affiliation{Yachay Tech University, School of Physical Sciences \& Nanotechnology, 100119 Urcuqu\'i, Ecuador}
\author{J.~Ovalle}
\email{jovalle@usb.ve}
\affiliation{Institute of Physics and Research Centre of Theoretical Physics and Astrophysics, Faculty of Philosophy and Science,
Silesian University in Opava, CZ-746 01 Opava, Czech Republic}
\affiliation{Departamento de F\'{\i}sica, Universidad Sim\'on Bol\'{\i}var, Apartado 89000, Caracas 1080A, Venezuela}
\author{A.~Sotomayor}
\email{adrian.sotomayor@uantof.cl}
\affiliation{Departamento de Matem\'aticas, Universidad de Antofagasta, Antofagasta, Chile.}
\author{Z.~Stuchlik}
\email{zdenek.stuchlik@fpf.slu.cz}
\affiliation{Institute of Physics and Research Centre of Theoretical Physics and Astrophysics, Faculty of Philosophy and Science,
Silesian University in Opava, CZ-746 01 Opava, Czech Republic}
\begin{abstract}
We employ the gravitational decoupling approach for static and spherically symmetric systems to develop a simple and powerful
method in order to a) continuously isotropize any anisotropic solution of the Einstein field equations,
and b) generate new solutions for self-gravitating distributions with the same or vanishing complexity factor. 
A few working examples are given for illustrative purposes.
\end{abstract} 
\maketitle
%
%
%
\section{Introduction}
\label{S:intro}
The Gravitational Decoupling (GD) was introduced in Ref.~\cite{Ovalle:2017fgl} 
as a systematic approach to study static and spherically symmetric self-gravitating systems governed by the 
Einstein field equations~\footnote{We shall use units with the speed of light $c=1$ and $k^2=8\,\pi\,G_{\rm N}$,
where $G_{\rm N}$ is Newton's constant.} 
\begin{equation}
\label{efe}
R_{\mu\nu}-\frac{1}{2}\,R\, g_{\mu\nu}
=
k^2\,\tilde{T}_{\mu\nu}
\ ,
\end{equation}
and containing (at least) two sources which only interact gravitationally. In its extended version, both time and radial components of the metric are affected and
these sources could exchange energy-momentum to provide the decoupling of Einstein's equations~\cite{Ovalle:2019qyi}.
The total energy-momentum tensor can thus be expressed as
\begin{equation}
\label{emt}
\tilde{T}_{\mu\nu}
=
T^{\rm}_{\mu\nu}
+
\alpha\,\theta_{\mu\nu}
\ ,
\end{equation}
where the constant $\alpha$ is here introduced for tracking the effects of $\theta_{\mu\nu}$ with respect to $T_{\mu\nu}$. 
As we will review briefly in the next Section, the key fact is that Eq.~\eqref{efe} can be split (continuously in $\alpha$)
into two sets of equations, one given by the Einstein field equations for the first source $T_{\mu\nu}$ (obtained in the limit $\alpha=0$)
and a set of ``quasi''-Einstein equations (proportional to $\alpha$) which describes the changes introduced by adding the second
source $\theta_{\mu\nu}$ (fully recovered for $\alpha=1$).
The way this split is implemented is by deforming the metric functions which solve the first set,
the deformation being then determined by the second set provided $\theta_{\mu\nu}$ is also given.  
\par
In fact, the GD is a generalization of the Minimal Geometric Deformation which was developed in
Refs.~\cite{Ovalle:2007bn,Ovalle:2009xk}
in the context of the Randall-Sundrum brane-world~\cite{Randall:1999ee,Randall:1999vf}, where the geometric deformation
is induced by the existence of extra spatial dimensions and $\alpha$ is naturally proportional to the inverse of the brane
tension~\cite{Casadio:2015gea,Ovalle:2015nfa,Casadio:2012pu,Ovalle:2013xla,Ovalle:2013vna,
Casadio:2013uma,Ovalle:2014uwa,Casadio:2015jva,Cavalcanti:2016mbe,Casadio:2016aum,daRocha:2017cxu,daRocha:2017lqj,
Fernandes-Silva:2017nec}~(for some resent applications see also~\cite{Casadio:2017sze,Fernandes-Silva:2018abr,
Fernandes-Silva:2019fez,daRocha:2019pla}). 
The main applications of this approach so far~\cite{Ovalle:2017wqi,Gabbanelli:2018bhs,Ovalle:2018umz,
Sharif:2018toc,Contreras:2018gzd,Sharif:2018pzr,Contreras:2018vph,Morales:2018nmq,Heras:2018cpz,Morales:2018urp,
Contreras:2018nfg,Panotopoulos:2018law,Ovalle:2018ans,Contreras:2019fbk,Maurya:2019wsk,Contreras:2019iwm,
Contreras:2019mhf,Gabbanelli:2019txr,Ovalle:2019lbs,Hensh:2019rtb,Cedeno:2019qkf,Leon:2019abq,Maurya:2019xcx,
Torres:2019mee,Rincon:2019jal}
were to build new solutions of Eq.~\eqref{efe} with $\alpha=1$ starting from known solutions generated by $T_{\mu\nu}$ alone
(that is, with $\alpha=0$).
In order to complete this construction, one needs to make assumptions about the second source, for instance
by fixing the equation of state for the tensor $\theta_{\mu\nu}$ (for the application of the GD beyond general relativity, see for instance Refs.~\cite{Sharif:2018tiz} and~\cite{Estrada:2019aeh}).
\par
In this paper we are instead interested in the different purpose of showing that the GD can be used to directly control specific
physical properties of a self-gravitating system.
For the sake of simplicity, we shall employ the Minimal Geometric Deformation (MGD) in which only the radial component
of the metric is modified and there is no direct exchange of energy between the two energy-momentum tensors
in Eq.~\eqref{emt}.
We shall then require that the complete system 
(for $\alpha=1$) enjoys specific properties, equal or different from those of the case $\alpha=0$.
In particular, we shall first require that the anisotropic pressure for $\alpha=0$ becomes isotropic for $\alpha=1$
in Section~\ref{s3} and impose conditions on the complexity factor which was recently introduced in Ref.~\cite{Herrera:2018bww}
in Section~\ref{s4}.
It is important to remark that the MGD does not involve any perturbative expansion and all results will be exact
for all values of $\alpha$.
Finally we summarise our conclusions in Section~\ref{con}.
%
%
%
%
\section{Gravitational decoupling of Einstein's equations} 
\label{s2}
We briefly review how the (M)GD works by starting from the standard Einstein field equations~\eqref{efe}
with two sources~\eqref{emt},
\begin{equation}
\label{corr2}
R_{\mu\nu}-\frac{1}{2}\,R\, g_{\mu\nu}
=
k^2\left({T}_{\mu\nu}+\alpha\,\theta_{\mu\nu}\right)
\ ,
\end{equation}
where the parameter $\alpha$ will be set to $0$ (respectively $1$) when we want to discard (fully include) the second source
$\theta_{\mu\nu}$.
Since the Einstein tensor in Eq.~\eqref{corr2} satisfies the Bianchi identity, the total source in
Eq.~(\ref{emt}) must be conserved, that is
\begin{equation}
\nabla_\mu\,\tilde{T}^{\mu\nu}=0
\ .
\label{dT0}
\end{equation}
\par 
For static spherically symmetric systems, the metric components $g_{\mu\nu}$ 
in Schwarzschild-like coordinates read 
\begin{equation}
ds^{2}
=
e^{\nu (r)}\,dt^{2}-e^{\lambda (r)}\,dr^{2}
-r^{2}\,d\Omega^2
\ ,
\label{metric}
\end{equation}
where $\nu =\nu (r)$ and $\lambda =\lambda (r)$ are functions of the areal
radius $r$ only and $d\Omega$ denotes the usual solid angle measure.
The metric~(\ref{metric}) must satisfy the Einstein equations~(\ref{corr2})
which, in terms of the two sources in~\eqref{emt}, explicitly read 
\begin{eqnarray}
\label{ec1}
k^2
\left(
T_0^{\ 0}+\alpha\,\theta_0^{\ 0}
\right)
&=&
\frac 1{r^2}
\left[
1-
{e^{-\lambda }}\left(1-r\,\lambda'\right)
\right]
\\
\label{ec2}
k^2
\left(T_1^{\ 1}+\alpha\,\theta_1^{\ 1}\right)
&=&
\frac 1{r^2}
\left[
1-
e^{-\lambda }\left(1+r\,\nu'\right)
\right]
\\
\label{ec3}
k^2
\left(T_2^{\ 2}+\alpha\,\theta_2^{\ 2}\right)
&=&
\frac {e^{-\lambda }}{4}
\left(
\lambda'\nu'
-2\,\nu''-\nu'^2
\right)
\nonumber
\\
&&
-
\frac {e^{-\lambda }}{2\,r}
\left(
\nu'-\lambda'
\right)
\ ,
\end{eqnarray}
where $f'\equiv \partial_r f$ and $\tilde{T}_3^{{\ 3}}=\tilde{T}_2^{\ 2}$ due to the spherical symmetry.
The conservation equation~(\ref{dT0}) is a linear combination of Eqs.~(\ref{ec1})-(\ref{ec3}) and reads
\begin{eqnarray}
0
&=&
\left(\tilde{T}_1^{\ 1}\right)'
-
\frac{\nu'}{2}\left(\tilde{T}_0^{\ 0}-\tilde{T}_1^{\ 1}\right)
-
\frac{2}{r}\left(\tilde{T}_2^{\ 2}-\tilde{T}_1^{\ 1}\right)
\nonumber
\\
&=&
\left({T}_1^{\ 1}\right)'
-
\frac{\nu'}{2}\left({T}_0^{\ 0}-{T}_1^{\ 1}\right)
-
\frac{2}{r}\left({T}_2^{\ 2}-{T}_1^{\ 1}\right)
\nonumber
\\
&&
+\alpha
\left[
\left({\theta}_1^{\ 1}\right)'
-
\frac{\nu'}{2}\left({\theta}_0^{\ 0}-{\theta}_1^{\ 1}\right)
-
\frac{2}{r}\left({\theta}_2^{\ 2}-{\theta}_1^{\ 1}\right)
\right]
\ .
\label{con11}
\end{eqnarray}
We can clearly identify in Eqs.~\eqref{ec1}-\eqref{ec3} an effective density  
\begin{equation}
\tilde{\rho}
=
T_0^{\ 0}
+\alpha\,\theta_0^{\ 0}
\equiv
\rho+\rho_\theta
\ ,
\label{efecden}
\end{equation}
an effective radial pressure
\begin{equation}
\tilde{p}_{r}
=
-T_1^{\ 1}-\alpha\,\theta_1^{\ 1}
\equiv
p_r
+
p_{\theta r}
\ ,
\label{efecprera}
\end{equation}
and an effective tangential pressure
\begin{equation}
\tilde{p}_{t}
=
-T_2^{\ 2}-\alpha\,\theta_2^{\ 2}
\equiv
p_t
+
p_{\theta t}
\ .
\label{efecpretan}
\end{equation}
These definitions clearly lead to the total anisotropy
\begin{equation}
\label{anisotropy}
\tilde{\Pi}
\equiv
\tilde{p}_{t}-\tilde{p}_{r}
\equiv
\Pi
+\Pi_\theta
\ ,
\end{equation}
where 
\begin{equation}
\Pi
=
p_t-p_r
\label{Pi}
\end{equation}
measures the anisotropy generated by the first source like $\Pi_\theta$ does for the second one.
\par
We will now implement the GD by considering a solution to Eqs.~(\ref{ec1})-(\ref{con11})
with $\alpha=0$, which we formally write as
\begin{equation}
\label{pfmetric}
ds^{2}
=
e^{\xi (r)}\,dt^{2}
-e^{\mu (r)}\,dr^{2}
-
r^{2}\,d\Omega^2
\ ,
\end{equation}
where 
\begin{equation}
\label{standardGR}
e^{-\mu(r)}
\equiv
1-\frac{k^2}{r}\int_0^r x^2\,T_0^{\, 0}(x)\, dx
=
1-\frac{2\,m(r)}{r}
\end{equation}
is the standard general relativistic expression containing the Misner-Sharp mass function $m=m(r)$.
The general effects of the second source $\theta_{\mu\nu}$ can then be encoded in the geometric deformation
undergone by the geometric functions $\xi \rightarrow \nu\,=\,\xi+\alpha\,g$ and
\begin{equation}
\label{gd2}
e^{-\mu} \rightarrow e^{-\lambda}=e^{-\mu}+\alpha\,f
\ . 
\end{equation}
From now on we just consider the simplest case of the MGD with a minimal deformation
$g(r) = 0$, hence only the radial metric component will be modified and $\nu=\xi$.
With the decomposition~(\ref{gd2}), the Einstein equations~(\ref{ec1})-(\ref{ec3})
split into two coupled sets:
i) the standard Einstein field equations for the energy-momentum tensor
$T_{\mu\nu}$ and metric~(\ref{pfmetric}),
\begin{eqnarray}
\label{ec1pf}
\rho
&=&
\frac{1}{k^2\,r^2}
\left[
1-
e^{-\mu }\left(1-r\,\mu'\right)
\right]
\\
\label{ec2pf}
p_r
&=&
-\frac{1}{k^2\,r^2}
\left[
1
-
e^{-\mu}\left( 1+r\,\xi'\right)
\right]
\\
\label{ec3pf}
p_t
&=&
-\frac {e^{-\mu }}{4\,k^2}
\left(\mu'\xi'-2\,\xi''-\xi'^2
-2\,\frac{\xi'-\mu'}r\right)
\ ,
\end{eqnarray}
with the conservation equation
\begin{equation}
\label{con111}
p_r'
+
\frac{\xi'}{2}\left(\rho+p_r\right)
=
\frac{2\,\Pi}{r}
\ ;
\end{equation}
and ii) the quasi-Einstein filed equations for the second source $\theta_{\mu\nu}$,
\begin{eqnarray}
\label{ec1d}
\rho_\theta
&=&
-\frac{\alpha\,f}{k^2\,r^2}
\left(1
+\frac{r\,f'}{f}
\right)
\\
\label{ec2d}
p_{\theta r}
&=&
\frac{\alpha\,f}{k^2\,r^2}
\left(1+r\,\xi'\right)
\\
\label{ec3d}
p_{\theta t}
&=&
\frac{\alpha\,f}{4\,k^2}
\left[
2\,\xi''+\xi'^2+2\frac{\xi'}{r}
+
\frac{f'}{f}
\left(\xi'+\frac{2}{r}\right)
\right]
\ ,
\end{eqnarray}
whose conservation equation likewise reads
\begin{eqnarray}
\label{con1112}
p_{\theta r}'
+
\frac{\xi'}{2}\left(\rho_\theta
+p_{\theta r}\right)
=
\frac{2\,\Pi_\theta}{r}
\ .
\end{eqnarray}
\section{Isotropization of compact sources}
\label{s3}
In the previous section, we noticed that the total anisotropy $\tilde\Pi$ in Eq.~\eqref{anisotropy}
can be different from the anisotropy $\Pi$ generated by the source $T_{\mu\nu}$.
We can therefore consider an {\em anisotropic\/} system~\eqref{ec1pf}-\eqref{con111} generated by 
$T_{\mu\nu}$ with $\Pi\not=0$ which is transformed into the {\em isotropic\/} system~\eqref{ec1}-\eqref{con11}
with $\tilde \Pi=0$ as a consequence of adding the source $\theta_{\mu\nu}$.
This change can be formally controlled by varying the parameter $\alpha$, with $\alpha=0$
representing the anisotropic system~\eqref{ec1pf}-\eqref{con111}, and $\alpha=1$
representing the isotropic system~\eqref{ec1}-\eqref{con11}, for which $\tilde\Pi=0$, or
\begin{equation}
\label{iso2}
\Pi_\theta
\equiv
{\theta}_1^{\ 1}
-
{\theta}_2^{\ 2}
=
-\Pi
\ .
\end{equation}
Replacing Eqs.~\eqref{ec2d} and \eqref{ec3d} in the condition~\eqref{iso2} yields a differential
equation for the geometric deformation in Eq.~\eqref{gd2}, namely
\begin{equation}
\label{iso3}
\frac{f'}{4\,k^2}\left(\xi'+\frac{2}{r}\right)
+\frac{f}{4\,k^2}\left(2\,\xi''+\xi'^2-\frac{2\,\xi'}{r}-\frac{4}{r^2}\right)
+\Pi
=0
\ .
\end{equation}
\par
As an example, we will implement the above approach in order to isotropize the compact self-gravitating
system sustained only by tangential stresses described by 
\begin{eqnarray}
\label{g00}
e^{\xi}
&=&
B^2
\left(1+\frac{r^2}{A^2}\right)
\ ,
\\
\label{g11}
e^{-\mu}
&=&
\frac{A^2+r^2}{A^2+3\,r^2}
\ ,
\\
\label{fden}
\rho
&=&
\frac{6 \left(A^2+r^2\right)}{k^2 \left(A^2+3\,r^2\right)^2}
\ ,
\\
\label{fpt}
p_t
&=&
\frac{3\,r^2}{k^2\,(A^2+3\,r^2)^2}
\ ,
\\
\label{fpr}
p_r
&=&
0
\ ,
\end{eqnarray}
where $0\le r \le R$ and $r=R$ defines the surface of the compact object.
A direct  interpretation of this class of solutions (albeit not unique, as pointed out in Ref.~\cite{Herrera:1997plx})
is in terms of a cluster of particles moving in randomly oriented circular orbits~\cite{Einstein:1939ms}.
The constants $A$ and $B$ can be determined from the matching conditions between this interior solution
and the exterior metric for $r>R$.
If we assume the exterior is the Schwarzschild vacuum,
\begin{eqnarray}
\label{ff1}
e^{\xi(R)}
&=&
1-\frac{2\,M}{R}
\\
\label{ff2}
e^{-\mu(R)}
&=&
1-\frac{2\,M}{R}
\\
\label{sff}
p_r(R)
&=&
0
\end{eqnarray}
are the necessary and sufficient conditions for a smooth matching of the two metrics.
This yields
\begin{equation}
\label{AB}
\frac{A^2}{R^2}
=
\frac{R-3\,M}{M}
\ ,
\qquad
B^2=1-\frac{3\,M}{R}
\ ,
\end{equation}
where $R>3\,M$ or $M/R<1/3$ in order to have $A^2>0$ and $B^2>0$.
\par
Plugging the solution~\eqref{g00}-\eqref{fpr} in the differential equation~\eqref{iso3}, we obtain the geometric
deformation 
\begin{eqnarray}
\label{f}
f(r)
=
\frac{r^2 \left(A^2+r^2\right)}{A^2+2\,r^2}
\left(\frac{1}{A^2+3\,r^2}
-\frac{1}{\ell^2}\right)
\ ,
\end{eqnarray}
where $\ell$ is an integration constant with dimensions of a length.
Using the metric functions~\eqref{gd2} and~\eqref{g00} in the field equation~\eqref{ec2},
the effective radial pressure in~\eqref{efecprera} is expressed as
\begin{equation}
\tilde{p}_r
=
\frac{\alpha\,f(r)\left(A^2+3\,r^2\right)}{k^2\,r^2 \left(A^2+r^2\right)}
\ .
\end{equation}
\begin{figure}[t]
	\center
	\includegraphics[scale=0.5]{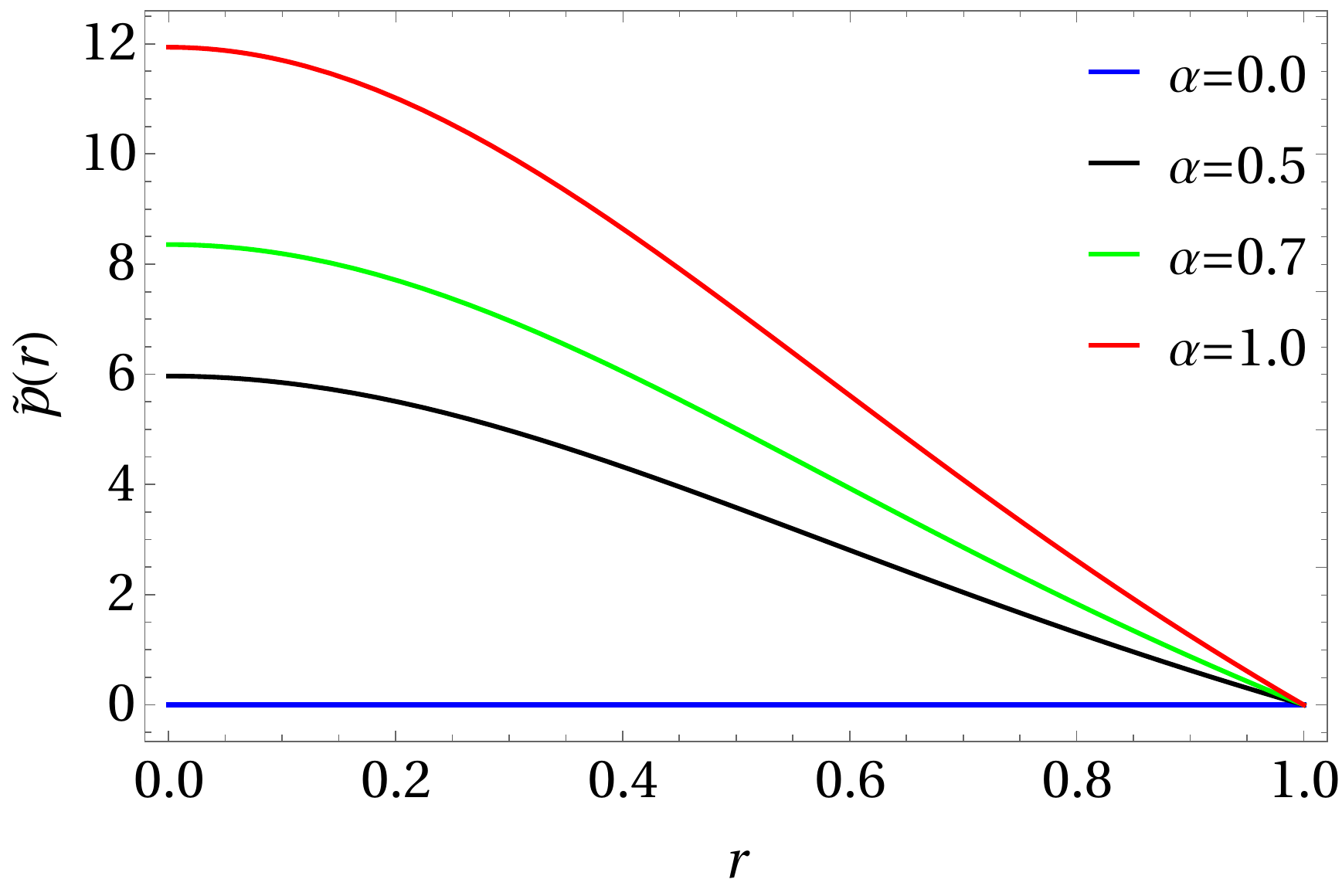}
	\\
	\caption{Isotropization: the radial pressure [$\tilde{p}\,\times\,10^3$] for different values of the
		parameter $\alpha$ for a distribution with compactness $M/R=0.2$.}
	\label{fig1}      
\end{figure}
Hence, the matching condition~\eqref{sff} for the outer Schwarzschild space-time yields
\begin{equation}
\label{fR}
f(R)=0\ ,
\end{equation}
which in turn leads to
\begin{equation}
\label{C}
\ell^2
=
{A^2+3\,R^2}
\ ,
\end{equation}
and the deformation takes the final form 
\begin{equation}
\label{ff}
f(r)=
\frac{3 \,r^2 \left(A^2+r^2\right)\left(R^2-r^2\right)}
{\left(A^2+2\, r^2\right)\left(A^2+3\,r^2\right)\left(A^2+3\, R^2\right)}
\end{equation}
Notice that the Misner-Sharp mass function $\tilde{m}$ of the system~\eqref{ec1}-\eqref{ec3}
is related with the mass function~\eqref{standardGR} of the system~\eqref{ec1pf}-\eqref{ec3pf}
by the simple expression
\begin{equation}
\frac{r}{2}\left(1-e^{-\lambda}\right)
\equiv
\tilde{m}(r)
=
m(r)-\frac{\alpha\,r\,f(r)}{2}
\ .
\label{tm}
\end{equation}
Hence, a direct consequence of~\eqref{fR} is that the total mass is the same for both cases, namely
\begin{equation}
\tilde{m}(R)=m(R)=M\ ,
\end{equation}
and therefore the values of the constants $A$ and $B$ remain the same as shown in~\eqref{AB}.
The deformation~\eqref{ff} generates an effective density
\begin{widetext}
\begin{equation}
\label{fden}
\tilde{\rho} (r,\alpha)
=
\rho(r)
-
\alpha\,\frac{18\,r^8-6\,r^6\,R^2+A^6\left(5\,r^2-3\,R^2\right)+2\,A^4\left(11\,r^4-5\,r^2\,R^2\right)+A^2\left(31\,r^6-9\,r^4\,R^2\right)}
{k^2\left(A^2+2\,r^2\right)^2\left(A^2+3\,r^2\right)^2\left(A^2+2\,R^2\right)}
\ ,
\end{equation}
\end{widetext}
an effective radial pressure
\begin{equation}
\label{fprd}
\tilde{p}_r(r,\alpha)
=
\frac{3\,\alpha  \left(R^2-r^2\right)}
{k^2\left(A^2+2 r^2\right) \left(A^2+3\,R^2\right)}\ ,
\end{equation}
and an effective tangential pressure $\tilde{p}_t=\tilde{p}_r+\tilde{\Pi}$
where the total anisotropy is given by
\begin{equation}
\label{fani}
\tilde{\Pi}(r,\alpha)=\frac{3\,(1-\alpha )\,r^2}{k^2 \left(A^2+3\,r^2\right)^2}
\ ,
\end{equation}
which vanishes, by construction, for $\alpha=1$.
\par
\begin{figure}[t]
	\center
	\includegraphics[scale=0.5]{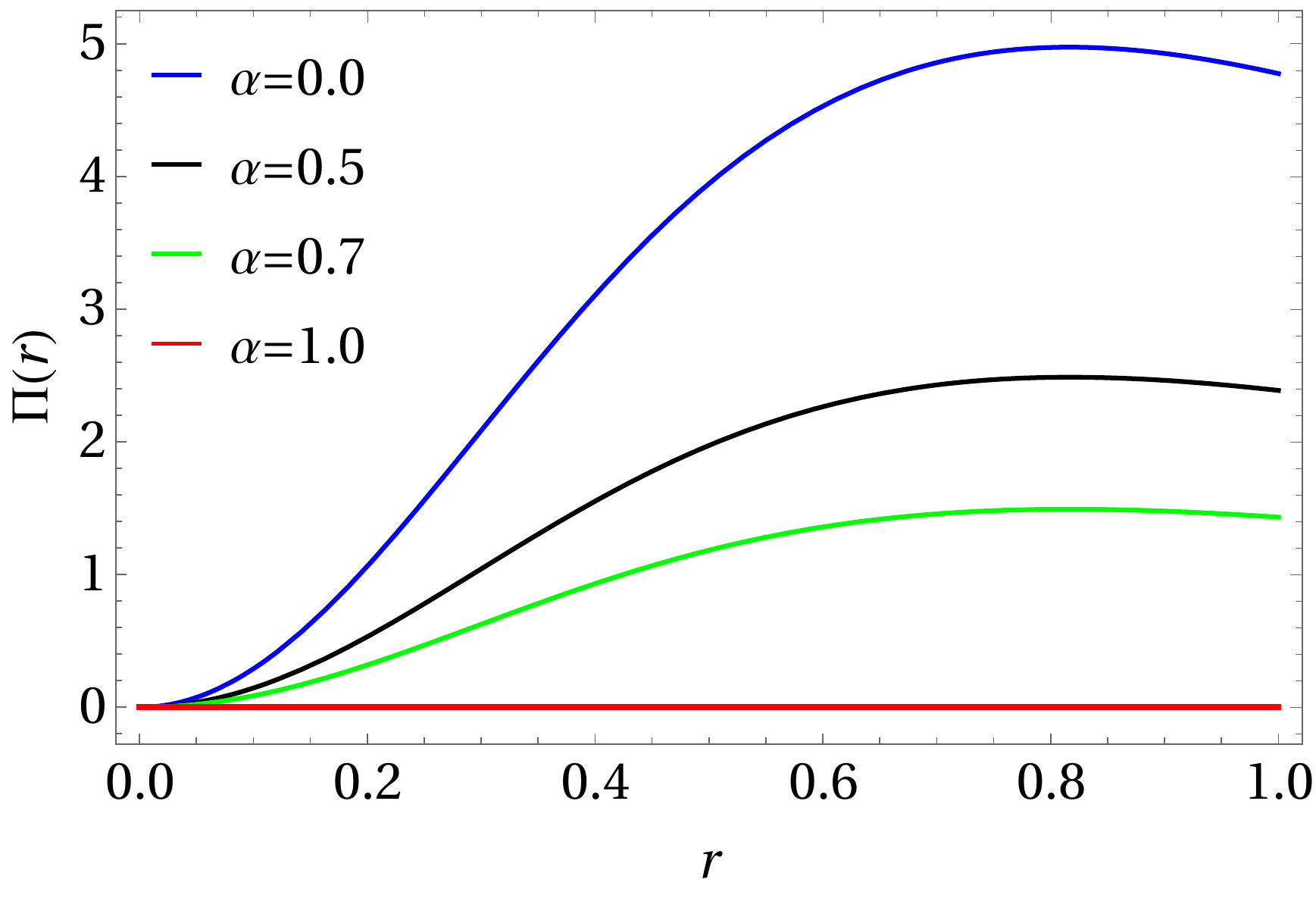}
	\\
	\caption{Isotropization: total anisotropy [$\Pi\,\times\,10^3$] for different values of the parameter $\alpha$
	for a distribution with compactness $M/R=0.2$.}
	\label{fig2}      
\end{figure}

The expressions~\eqref{g00},~\eqref{tm} and~\eqref{fden}-\eqref{fani} are exact solutions of the
Einstein field equations~\eqref{ec1}-\eqref{ec3} for all values of $\alpha$.
We can further see that the case $\alpha=0$ represents the anisotropic model in~\eqref{g00}-\eqref{fpr},
which is continuously deformed into the isotropic case represented by $\alpha=1$.
Hence we can follow in details the isotropization process by continuously varying the parameter $\alpha$
between these two values [see Fig~\ref{fig1} and~\ref{fig2}, where the effective pressure in Eq.~\eqref{fprd}
and the anisotropy in Eq.~\eqref{fani} are shown for a few values of $\alpha$].
%
%
%
%
%
%
%
%
\section{Complexity of compact sources}
\label{s4}
The notion of complexity for static and spherically symmetric self-gravitating systems we are interested in here
was introduced recently  in Ref.~\cite{Herrera:2018bww}, and further extended to the dynamical case
in Ref.~\cite{Herrera:2018czt}
(for some applications, see {\rm e.g.}~Refs.~\cite{Abbas:2018idr,Sharif:2018pgq}).
The main characteristic of this notion is that it assigns a zero value of the complexity factor to uniform and
isotropic distributions (the least complex system).
\par
The complexity of a given static and spherically symmetric self-gravitating system is measured 
by the complexity factor $Y_{\rm TF}$, which is a scalar function defined in terms of the anisotropy $\Pi$
and gradient $\rho'$ of the energy-density as~\cite{Herrera:2018bww}
\begin{equation}
\label{YTF}
Y_{\rm TF}(r)=k^2\,\Pi(r)-\frac{k^2}{2\,r^3}\int^r_0x^3\rho'(x)\,dx\ .
\end{equation}
It describes the influence of these two functions  on the Tolman mass $m_{\rm T}$
which, for the same distribution of matter, is defined as 
\begin{equation}
\label{mtd}
m_{\rm T}(r)
=
\frac{k^2}{2}\int_0^r\,e^{(\xi+\lambda)/2}
\left(\rho+p_r+2\,p_t\right)
x^2\,dx
\ .
\end{equation}
The above definition gives the energy contained inside a fluid sphere of radius $r$, and it has a clear
physical interpretation.
In fact, we recall that we can write the Tolman mass
as a function of the metric components in Eq.~\eqref{metric} as
\begin{eqnarray}
\label{mt}
m_{\rm T}
=
\frac{r^2\,\xi'}{2}\,e^{(\xi-\lambda)/2}
\end{eqnarray}
and that the gravitational acceleration of
a test particle, instantaneously at rest in the  static gravitational field~\eqref{metric}, is given by
\begin{eqnarray}
\label{a}
a=-\frac{e^{-\xi/2}\,m_{\rm T}}{r^2}
\ ,
\end{eqnarray}
which shows that $m_{\rm T}$ is the active gravitational mass (for more details,
see Refs.~\cite{Herrera:1997plx,Herrera:1997si}). 
\par
In terms of the complexity factor, we can write the Tolman mass as
\begin{equation}
\label{tolmanmass}
m_{\rm T}
=
M_{\rm T}\left(\frac{r}{R}\right)^3
+r^3\int^R_r\frac{e^{(\xi+\lambda)/2}}{x}\,Y_{\rm TF}\,dx
\ ,
\end{equation}
where $M_{\rm T}$ represents the total Tolman mass of an isotropic and uniform stellar system of the same
radius $R$.
Hence, we see that $Y_{\rm TF}$ can quantify the departure of the Tolman mass $m_{\rm T}$ of a given system
from the Tolman mass $M_{\rm T}$ of a uniform isotropic fluid when the anisotropy and density gradient do not vanish.
It is in fact clear from Eq.~\eqref{YTF} that a uniform isotropic fluid will have zero complexity factor, but this does
not mean that a stellar configuration with vanishing complexity factor is uniform and isotropic.
\par
From Eq.~\eqref{YTF} we see that the complexity factor for the system~\eqref{ec1}-\eqref{ec3} takes the form
\begin{eqnarray}
\tilde{Y}_{\rm TF}
&=&
k^2\,\tilde{\Pi}-\frac{k^2}{2\,r^3}\int^r_0\tilde{r}^3\tilde{\rho}'\,d\tilde{r}
\nonumber
\\
\label{cftilde2}
&=&k^2\,{\Pi}-\frac{k^2}{2\,r^3}\int^r_0\tilde{r}^3{\rho}'\,d\tilde{r}
\nonumber
\\
&&
+
k^2\,{\Pi}_\theta-\frac{k^2}{2\,r^3}\int^r_0\tilde{r}^3{\rho_\theta}'\,d\tilde{r}
\ ,
\end{eqnarray}
which we can write as
\begin{equation}
\label{addition}
\tilde{Y}_{\rm TF}={Y}_{\rm TF}+{Y}^{\theta}_{\rm TF}
\ ,
\end{equation}
where ${Y}_{\rm TF}$ and ${Y}^{\theta}_{\rm TF}$ are the complexity factors corresponding
to the systems~\eqref{ec1pf}-\eqref{ec3pf} and~\eqref{ec1d}-\eqref{ec3d} respectively.
We conclude that the complexity factor is an additive quantity.
Hence, the complexity factor of a gravitational system formed by two coexisting gravitational
sources, $T^{\rm}_{\mu\nu}$ and $\theta_{\mu\nu}$, will be the sum of the complexity factors
of the two sources.
We remark that this result is independent of the MGD described in Section~\ref{s2},
but it implies that we can employ the GD in order to relate two different systems with the same
or different complexity factors.
\subsection{Two systems with the same complexity factor}
\label{SS:comp1}
We first consider a case in which the complexity factor ${Y}_{\rm TF}$ associated with the
energy-momentum tensor $T_{\mu\nu}$ remains invariant after we add the second source
$\theta_{\mu\nu}$, that is $\tilde{Y}_{\rm TF}={Y}_{\rm TF}$, and therefore
${Y}^{\theta}_{\rm TF}=0$ or
\begin{equation}
\label{y1}
\Pi_\theta
=
\frac{1}{2\,r^3}\int^r_0\tilde{r}^3\rho_\theta'\,d\tilde{r}
\ .
\end{equation}
Using Eqs.~\eqref{ec1d}-\eqref{ec3d} yields
\begin{equation}
\frac{1}{2\,r^3}\int^r_0\tilde{r}^3\rho_\theta'\,d\tilde{r}
=
\frac{\alpha}{k^2}
\left(\frac{f}{r^2}-\frac{f'}{2\,r}\right)
\end{equation}
and the condition~\eqref{y1} becomes the first order differential equation
\begin{eqnarray}
f' \left(\xi'+\frac{4}{r}\right)
+f
\left(2\,\xi''+\xi'^2-\frac{2\, \xi'}{r}-\frac{8}{r^2}\right)
=0
\ .
\label{eqf}
\end{eqnarray}
Any solution of Eq.~\eqref{eqf} can be used to determine the source $\theta_{\mu\nu}$ through~\eqref{ec1d}-\eqref{ec3d}.
In other words, given a solution with metric functions $\xi$ and $\mu$ for the Einstein field equations~\eqref{ec1pf}-\eqref{ec3pf},
we can find a second solution to~\eqref{ec1}-\eqref{ec3} with the same complexity factor by imposing the condition~\eqref{y1}.
Like with isotropization in Section~\ref{s3}, the parameter $\alpha$ can be implemented to continuously follow this process
by identifying the original solution with the case $\alpha=0$ and the final solution with $\alpha=1$.
However, since Eq.~\eqref{eqf} does not contain $\alpha$, we can now actually require that the complexity remains
the same for all values of $\alpha$.
By implementing this procedure we will moreover find that the matching conditions~\eqref{ff1}-\eqref{sff} play a fundamental
role in the determination of the final result in that the condition $\tilde{Y}_{\rm TF}={Y}_{\rm TF}$ can only be satisfied if
we change the compactness of the system.
\par
 Let us start by considering as solution to~\eqref{ec1pf}-\eqref{ec3pf} the Tolman~IV metric for
 perfect fluids~\cite{Tolman:1939jz},
 \begin{eqnarray}
 \label{tolman00}
 e^{\xi}
 &=&
 B^2
 \left(1+\frac{r^2}{A^2}\right)
 \\
 \label{tolman11}
 e^{-\mu}
 &=&
 \frac{\left(C^2-r^2\right)\left(A^2+r^2\right)}{C^2\left(A^2+2\,r^2\right)}
 \ ,
 \end{eqnarray}
 which is generated by the density
 \begin{equation} 
 \label{tolmandensity}
 \rho
=
 \frac{3\,A^4+A^2\left(3\,C^2+7\,r^2\right)+2\, r^2 \left(C^2+3 \,r^2\right)}{k^2\,C^2\left(A^2+2\,r^2\right)^2}
 \ ,
 \end{equation}
 and isotropic pressure
 \begin{equation}
 \label{tolmanpressure}
 p
 =
 \frac{C^2-A^2-3\,r^2}{k^2\,C^2\left(A^2+2\,r^2\right)}
 \ .
 \end{equation}
 The constants $A$, $B$ and $C$ are again determined from the
 matching conditions~\eqref{ff1}-\eqref{sff}, which yield the same
 values~\eqref{AB} and
 \begin{equation}
 \frac{C^2}{R^2}
 =
 \frac{R}{M}
 \ .
 \end{equation}
From the definition~\eqref{YTF} we obtain the complexity factor 
\begin{equation}
\label{Ytolman}
Y_{\rm TF}
=
\frac{\left(A^2+2\, C^2\right)r^2}
{C^2\left(A^2+2\, r^2\right)^2}
\ .
\end{equation} 
From the metric function~\eqref{tolman00}, we can then compute the deformation which keeps this factor
unchanged by solving Eq.~\eqref{eqf}, and obtain
\begin{equation}
f(r)
=
\frac{r^2 \left(A^2+r^2\right)}{\ell^2\left(2 \,A^2+3\,r^2\right)}
\ ,
\end{equation} 
where $\ell$ is an integration constant (with dimensions of length).
According to~\eqref{gd2}, the new radial metric component therefore reads
\begin{equation}
\label{g11fa}
 e^{-\lambda}
 =
 \frac{\left(C^2-r^2\right)\left(A^2+r^2\right)}{C^2\left(A^2+{2\,r^2}\right)}
 +\frac{\alpha\,r^2 \left(A^2+r^2\right)}{\ell^2\left(2 \,A^2+3\,r^2\right)}
 \end{equation}
and generates an effective density
\begin{equation}
\label{faden}
\tilde{\rho}(r,\alpha,\ell)
=
\rho(r)
-\frac{\alpha\left(6\, A^4+13\, A^2\, r^2+9\,r^4\right)}{\ell^2\,k^2 \left(2\, A^2+3\, r^2\right)^2}
\ ,
\end{equation}
an effective radial pressure
\begin{equation}
\label{fapr}
\tilde{p}_r(r,\alpha,\ell)
=
p(r)
+\frac{\alpha\left(A^2+3\, r^2\right)}
{\ell^2\,k^2\left(2 \,A^2+3 \,r^2\right)}
\ ,
\end{equation}
and an effective tangential pressure $\tilde{p}_t=\tilde{p}_r+\tilde{\Pi}$
where the anisotropy is given by
\begin{equation}
\label{faani}
\tilde{\Pi}(r,\alpha,\ell)
=
\frac{\alpha\,  A^2\, r^2}
{\ell^2\,k^2 \left(2\, A^2+3\,r^2\right)^2}
\ .
\end{equation}
\par
\begin{figure}[t]
\center
\includegraphics[scale=0.5]{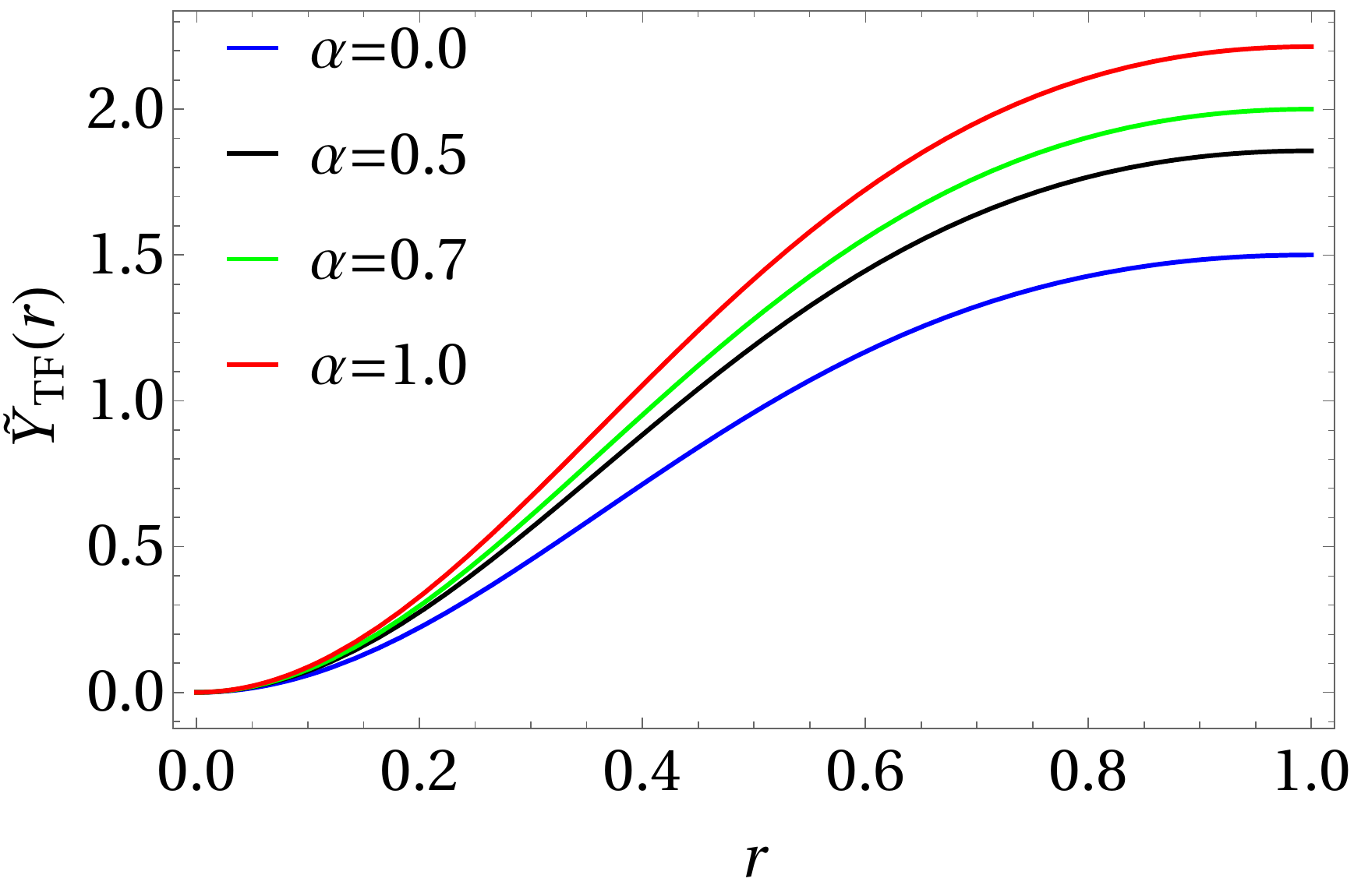}
\\
\caption{Complexity factor [$\tilde{Y}_{\rm TF}\times 10$] for different values of
the parameter $\alpha$ starting from the Tolman IV isotropic solution ($\alpha=0$).}
\label{fig4}      
\end{figure}
The expressions~\eqref{tolman00} and~\eqref{g11fa}-\eqref{faani} describe an exact solution of the
Einstein field equations~\eqref{ec1}-\eqref{ec3}.
This is a new  anisotropic version of the Tolam IV solution~\eqref{tolman00}-\eqref{tolmanpressure},
whose complexity factor $\tilde{Y}_{\rm TF}$ is formally the same as that in Eq.~\eqref{Ytolman}.
However, after imposing the matching conditions~\eqref{ff1}-\eqref{sff} to determinate the new values
for $A$, $B$ and $C$ in the solution~\eqref{tolman00} and~\eqref{g11fa}-\eqref{faani}, we find that
$A$ and $B$ have the same expressions as shown in Eq.~\eqref{AB}, but $C$ is promoted to a function
of the anisotropic parameter $\alpha$ (and the length $\ell$), namely
\begin{equation}
C^2_{\alpha\ell}
=
\frac{R^3}{M}
-\frac{\alpha  \left(A^2+2\, R^2\right)\left(A^2+3 \,R^2\right)^2}
{\alpha\left(A^4+5\,A^2\,R^2+6\,R^4\right)
+\ell^2\left(2\,A^2+3\,R^2\right)}
\ ,
\end{equation}
and the complexity factor becomes 
\begin{equation}
\label{Ytolman2}
\tilde{Y}_{\rm TF}(r,\alpha,\ell)
=
\frac{\left[A^2+2\, C^2_{\alpha\ell}\right]r^2}{C^2_{\alpha\ell}\left(A^2+2 r^2\right)^2}
\ .
\end{equation} 
Comparing the expressions~\eqref{Ytolman} and~\eqref{Ytolman2} shows that varying $\alpha$
in fact changes the complexity factor (see Fig.~\ref{fig4}) unless we also change the mass $M\to M_{\alpha\ell}$ and
the radius $R\to R_{\alpha\ell}$ in such a way that
\begin{equation}
C_{\alpha\ell}(M_{\alpha\ell},R_{\alpha\ell})
=
C(M,R)
=
\frac{R^3}{M}
\ .
\end{equation}
In the above equation for $M_{\alpha\ell}$ and $R_{\alpha\ell}$, we can set $\alpha=1$ without loss of generality,
but we are still left with the freedom to set the arbitrary length scale $\ell$.
This means that we can generate a continuous family of systems with different mass $M_\ell$ and radius $R_\ell$
but the same total complexity factor $Y_{\rm TF}$ in Eq.~\eqref{Ytolman}. 
\subsection{Generating solutions with zero complexity}
\label{SS:comp2}
We will now show how one can build a solution with $\tilde{Y}_{\rm TF}=0$ starting
from a first solution with ${Y}_{\rm TF}\neq 0$.
According to Eq.~\eqref{addition}, we can therefore require the condition
\begin{eqnarray}
\label{y2}
\tilde Y_{\rm TF}
=
Y_{\rm TF}
+
k^2\,{\Pi}_\theta
-\frac{k^2}{2\,r^3}
\int^r_0\tilde{r}^3{\rho_\theta}'\,d\tilde{r}=0
\ ,
\end{eqnarray}
for $\alpha=1$~\footnote{We recall that $\Pi_\theta$ in Eq.~\eqref{anisotropy} and $\rho_\theta$ in Eq.~\eqref{efecden}
are both proportional to $\alpha$, so that they vanish for $\alpha=0$ by construction.}.
%
%
%
%
%
%
%
%
%
%
Using~Eqs.~\eqref{ec1d}-\eqref{ec3d} in the condition~\eqref{y2}, we obtain the first order differential equation
for the geometric deformation
\begin{equation}
\frac{f'}{4} \left(\xi'+\frac{4}{r}\right)
+\frac{f}{4} \left(2 \,\xi''+\xi'^2-\frac{2\,\xi'}{r}-\frac{8}{r^2}\right)
+{Y_{\rm TF}}
=0
\ ,
\label{eqC}
\end{equation}
whose solution can be used to generate a system with vanishing complexity factor $\tilde{Y}_{\rm TF}=0$
for $\alpha=1$. 
%
%
 %
 %
 %
 \begin{figure}[t]
 	\center
 	\includegraphics[scale=0.5]{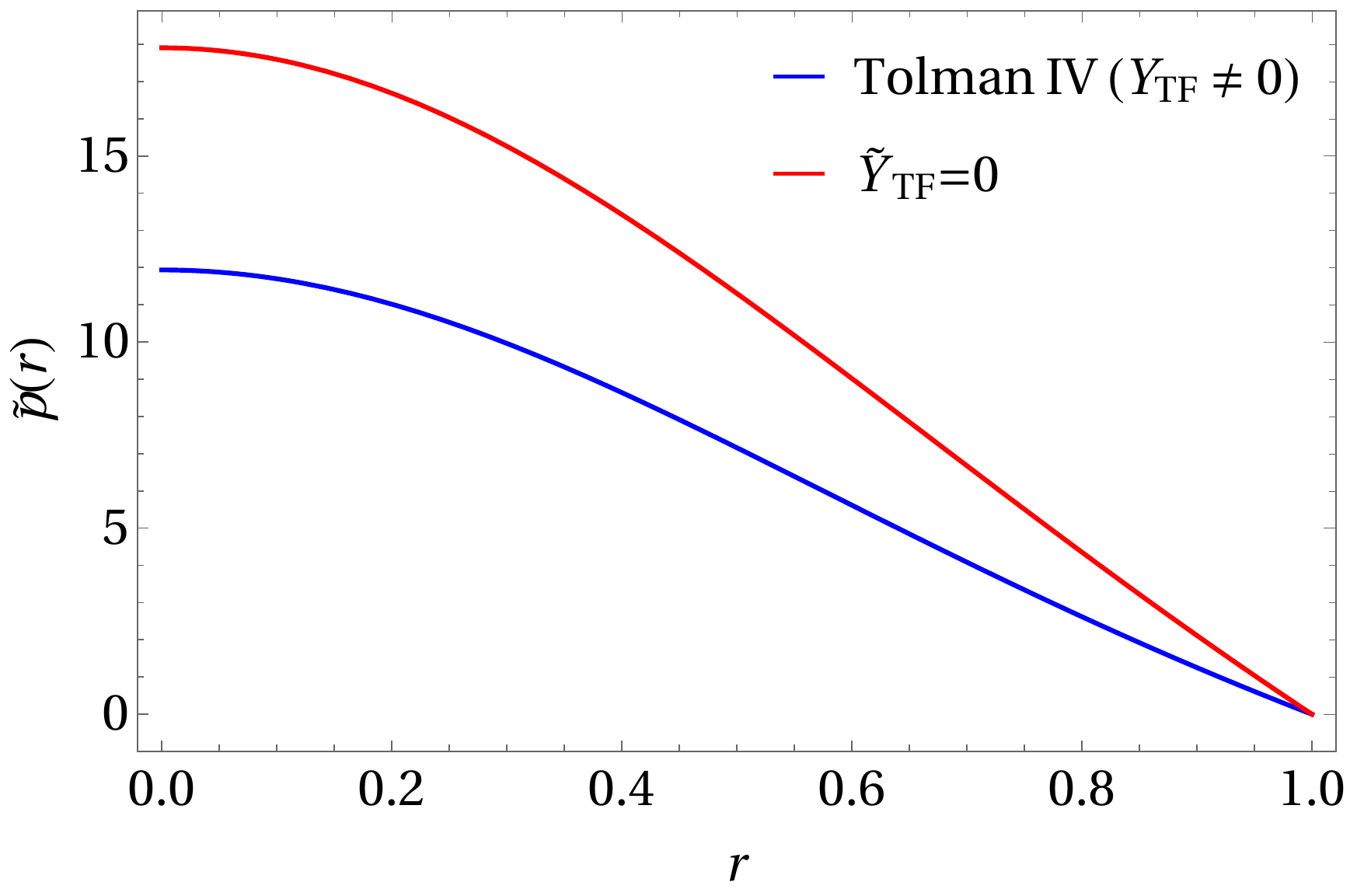}
 	\\
 	\caption{Radial pressure [$\tilde{p}\times 10^3$] for the Tolman IV solution ($Y_{\rm TF}\neq 0$)
	and its anisotropic version with $\tilde{Y}_{\rm TF}=0$ for a distribution with compactness $M/R=0.2$.}
 	\label{fig6}      
 \end{figure}
\par
Let us consider again the Tolman IV solution~\eqref{tolman00}-\eqref{tolmanpressure}.
Using the metric function~\eqref{tolman00} and the complexity factor~\eqref{Ytolman}, Eq.~\eqref{eqC}
can be solved exactly to yield
\begin{equation}
f
=
\frac{r^2 \left(A^2+r^2\right)}{\ell^2\left(2\,A^2+3\,r^2\right)}
\left[
1
+
\frac{\ell^2\left(A^2+2\,C^2\right)}
{2 \,C^2\left(A^2+2 r^2\right)}
\right]
\ ,
\end{equation}
where $\ell$ is an arbitrary integration constant with dimensions of a length.
The corresponding new radial metric component 
will change the complexity factor in Eq.~\eqref{Ytolman} to
\begin{equation}
\label{yd}
\tilde{Y}_{\rm TF}(r,\alpha)
=
(1-\alpha )\,
\frac{\left(A^2+2\,C^2\right)r^2}{C^2 \left(A^2+2\, r^2\right)^2}
\ ,
\end{equation}
which precisely interpolates continuously between the original value (for $\alpha=0$) and 
vanishing complexity (for $\alpha=1$).
Note that the arbitrary scale $\ell$ does not appear explicitly in Eq.~\eqref{yd}.
However, it will affect the value of the constant $C$ via the matching conditions
like in the previous case.
\par
For $\alpha=1$, we can determine all relevant quantities explicitly.
The matching conditions~\eqref{ff1}-\eqref{sff} with the outer Schwarzschild vacuum
yield the same $A$ and $B$ shown in Eq.~\eqref{AB}, while $C$ is now given by
\begin{equation}
C^2
=
\frac{3\,\ell^2 \left(A^2+3\, R^2\right)}{2\left(A^2+3\, R^2+3\,\ell^2\right)}
\ .
\end{equation}
The radial metric component then takes the final form
\begin{equation}
\label{g11fc}
e^{-\lambda}
=
\frac{\left(A^2+r^2\right) \left(2\,A^2-3\, r^2+6\, R^2\right)}
{\left(2\,A^2+3\,r^2\right) \left(A^2+3\,	R^2\right)}
\ ,
\end{equation}
the effective radial pressure reads (see also Fig.~\ref{fig6})
\begin{equation}
\label{fcpr}
\tilde{p}_r
=
\frac{9 \left(R^2-r^2\right)}
{k^2\left(2\, A^2+3\, r^2\right)
\left(A^2+3 R^2\right)}
\ ,
 \end{equation}
the effective density is
\begin{equation}
\label{fbden}
\tilde{\rho}
=
\frac{3 \left[8\, A^4+2\, A^2 \left(7\,r^2+3\, R^2\right)+3\, r^2 \left(3\,r^2+R^2\right)\right]}
{k^2\left(2\, A^2+3\, r^2\right)^2	\left(A^2+3 R^2\right)}
\ ,
 \end{equation}
and the effective tangential pressure $\tilde{p}_t=\tilde{p}_r+\tilde{\Pi}$,
where the anisotropy is given by
\begin{equation}
\label{fbani}
\tilde{\Pi}
=
-\frac{3\, r^2 \left(2\, A^2+3\,R^2\right)}
{k^2 \left(2\, A^2+3\,r^2\right)^2 \left(A^2+3\,R^2\right)}
\ .
\end{equation}
The main difference with respect to the case in Section~\ref{SS:comp1} is that
the complexity~\eqref{yd} vanishes for $\alpha=1$ regardless of $M$ and $R$, and therefore
for any values of $\ell$:
we have mapped the Tolman~IV fluid of given mass $M$, radius $R$ and complexity~\eqref{Ytolman}
into a whole family of systems with the same mass $M$ and radius $R$ but vanishing complexity
parametrized by the length scale $\ell$.
\section{Conclusions}
\label{con}
The GD approach is a very effective way to investigate self-gravitating systems with sources
described by more than one (spherically symmetric) energy-momentum tensor.
Given an exact solution generated by one of such sources, it will allow one to obtain exact solutions with more
sources.
In most of the previous papers, new solutions were obtained by assuming particular equations of state for the added
energy-momentum tensors, or field equations for the added matter sources.
In this work we have instead considered the different task of employing the GD in order to
impose specific physical properties satisfied by the whole system.
\par
In order to keep the presentation simpler, we just considered two energy-momentum tensors and the MGD
in which only the radial component of the metric is modified,
although the approach can be straightforwardly generalised to more sources and to the GD in which the time
component of the metric is deformed as well.
The specific properties we required were isotropic pressure starting from the anisotropic solution~\eqref{g00}-\eqref{fpr}
and control over the complexity factor starting from the Tolman IV solution~\eqref{tolman00}-\eqref{tolmanpressure}.
The examples we provided are mostly meant to illustrate the flexibility and effectiveness of our procedure and different
physical requirements could indeed be demanded.
\section{Acknowledgements}
R.C.~is partially supported by the INFN grant FLAG and his work has also been carried out in the framework of activities
of the National Group of Mathematical Physics (GNFM, INdAM) and COST action {em Cantata\/}. J.O. thanks Luis Herrera for useful discussion and comments. J.O. acknowledges the support of the Institute of Physics and Research Centre of Theoretical Physics and Astrophysics, at the Silesian University in Opava. S.Z.~has been supported by the Czech Science Agency Grant No.19-03950S.
\section*{References}
\bibliography{references}
\bibliographystyle{iopart-num.bst}
\end{document}